\documentclass{emulateapj}
\usepackage{amsmath,natbib}

\newcommand\msun{\hbox{{$M_{\odot}$}}}

\newcommand\asca{{\sl ASCA}}

\newcommand\chandra{{\sl Chandra}}
\newcommand\xmm{{\sl XMM}}

\newcommand\kmsmpc{{\rm km s$^{-1}$ Mpc$^{-1}$}}

\newcommand\lx{\hbox{{$L_{\rm x}$}}}

\newcommand\rvir{{\hbox{$r_{\rm vir}$}}}
\newcommand\mvir{{\hbox{$M_{\rm vir}$}}}

\newcommand\omegam{\hbox{{$\Omega_{\rm m}$}}}
\newcommand\omegalambda{\hbox{{$\Omega_{\Lambda}$}}}

\newcommand\lcdm{\hbox{{$\Lambda$CDM}}}
\newcommand\lone{\hbox{{$\Lambda$CDM1}}}
\newcommand\lthree{\hbox{{$\Lambda$CDM3}}}
\newcommand\logten{\hbox{{$\log_{10}$}}}
\newcommand\sige{\hbox{{$\sigma_8$}}}

\begin{document} 

\title{The X-Ray Concentration-Virial Mass Relation}
\author{David A. Buote\altaffilmark{1}, Fabio Gastaldello\altaffilmark{1}, 
Philip J. Humphrey\altaffilmark{1}, Luca Zappacosta\altaffilmark{1},
James S. Bullock\altaffilmark{1}, \\ Fabrizio
Brighenti\altaffilmark{2,3}, \& William G. Mathews\altaffilmark{2}}
\altaffiltext{1}{Department of Physics and Astronomy, University of
California at Irvine, 4129 Frederick Reines Hall, Irvine, CA 92697-4575}
\altaffiltext{2}{UCO/Lick Observatory, Board of Studies in Astronomy
and Astrophysics, University of California, Santa Cruz, CA 95064}
\altaffiltext{3}{Dipartimento di Astronomia, Universit\`a di Bologna,
via Ranzani 1, Bologna 40127, Italy}

\slugcomment{Accepted for Publication in The Astrophysical Journal}

\begin{abstract}

We present the concentration ($c$)-virial mass ($M$) relation of 39
galaxy systems ranging in mass from individual early-type galaxies up
to the most massive galaxy clusters, $(0.06-20) \times
10^{14}\msun$. We selected for analysis the most relaxed systems
possessing the highest quality data currently available in the
\chandra\ and \xmm\ public data archives. A power-law model fitted to
the X-ray $c-M$ relation requires at high significance ($6.6\sigma$)
that $c$ decreases with increasing $M$, which is a general feature of
CDM models. The median and scatter of the $c-M$ relation produced by
the flat, concordance \lcdm\ model ($\omegam=0.3$, $\sige=0.9$) agrees
with the X-ray data provided the sample is comprised of the most
relaxed, early forming systems, which is consistent with our selection
criteria. Holding the rest of the cosmological parameters fixed to
those in the concordance model the $c-M$ relation requires
$0.76<\sige<1.07$ (99\% conf.), assuming a 10\% upward bias in the
concentrations for early forming systems. The tilted, low-$\sige$
model suggested by a new WMAP analysis is rejected at $99.99\%$
confidence, but a model with the same tilt and normalization can be
reconciled with the X-ray data by increasing the dark energy equation
of state parameter to $w\approx -0.8$.  When imposing the additional
constraint of the tight relation between \sige\ and \omegam\ from
studies of cluster abundances, the X-ray $c-M$ relation excludes
($>99\%$ conf.) both open CDM models and flat CDM models with
$\omegam\approx 1$. This result provides novel evidence for a flat,
low-$\omegam$ universe with dark energy using observations only in the
local ($z\ll 1$) universe.  Possible systematic errors in the X-ray
mass measurements of a magnitude $\approx 10\%$ suggested by CDM
simulations do not change our conclusions. We discuss other sources of
systematic error in the measurements and theoretical predictions that
need to be addressed for future precision cosmological studies using
the $c-M$ relation.
\end{abstract}


\section{Introduction}
\label{intro}

The current cosmological paradigm of a flat universe with low matter
density (\omegam) consisting of Cold Dark Matter (CDM) and dark energy
has become established recently through a variety of astronomical
observations, especially the Cosmic Microwave Background (CMB),
supernova Hubble diagram, and statistics of large galaxy surveys
\citep[e.g.,][]{sper03a,asti06a,sper06a}. These key techniques probe the
large-scale, high-redshift universe where the development of cosmic
structure may be studied in the linear clustering regime ($\delta\ll
1$). Although the standard model appears to have withstood every test
thus far \citep[e.g.,][]{prim06a}, testing the \lcdm\ model on small
scales, fully into the non-linear regime ($\delta\gg 1$), will further
our understanding of galaxy formation and evolution and may also
refine the fundamental parameters of the world model itself.  

The dark matter halos of galaxies and clusters are the largest
virialized manifestations of small-scale, non-linear clustering.  An
important probe of the structure of dark matter halos is the
relationship between concentration and virial mass.  The concentration
parameter is defined as, $c\equiv r_{\Delta}/r_s$, where $r_{\Delta}$
-- the ``virial radius'' -- is usually taken to be the radius within
which the average density equals $\Delta\rho_c$, where $\rho_c$ is the
critical density of the universe, and $\Delta$ is a number typically
between $100-500$. The quantity $r_s$ is the scale radius of the NFW
profile \citep{nfw}, but it is replaced by $r_{-2}$, the radius where
the logarithmic density slope equals -2, for more general profiles
\citep[][]{nava04a,grah06a}. The virial mass ($M$) is the mass
enclosed within $r_{\Delta}$.

N-body simulations of CDM models find that $c$ declines slowly with
increasing $M$ but with substantial intrinsic scatter, independent of
$M$ (\citealt{bull01a}, hereafter B01; \citealt{dola04a}, hereafter
D04; \citealt{kuhl05a,shaw06a,macc06a}). For a given $M$ the value of
$c$ in CDM models varies significantly with changes in cosmological
parameters, particularly the normalization \sige, \omegam, and $w$,
the parameter representing the dark energy equation of state
(\citealt{eke01a,alam02a}; D04; \citealt{kuhl05a}). The amount of
intrinsic scatter in the $c-M$ relation is a robust prediction of CDM
models, though preferentially selecting relaxed, early forming systems
yields smaller scatter than for the entire halo population
\citep{jing00a,wech02a,macc06a}.

The expected decrease of $c$ with increasing $M$ in CDM models has yet
to be confirmed by observations. Optical studies of groups and
clusters ($M>10^{13}\msun$) using several different techniques --
galaxy dynamics \citep{bivi06a,loka06a}, redshift-space caustics
\citep{rine06a}, and weak gravitational lensing \citep{mand06a} -- all
are consistent with no variation in $c$. However, the $c-M$ relations
obtained by these studies are consistent with the concordance \lcdm\
model.

X-ray studies of the $c-M$ relation using data from the \asca\
satellite \citep{wu00a,sato00a} found, $c\sim M_{200}^{-0.5}$, much
steeper than produced in CDM models (B01; D04; \citealt{macc06a}).
However, it has only become possible with data from the \chandra\ and
\xmm\ satellites to obtain spatially resolved temperature measurements
of sufficient quality for reliable X-ray constraints on the mass
profiles of elliptical galaxies, galaxy groups, and clusters
\citep[e.g., see reviews by][and references therein]{buot03d,arna05a}.
Indeed, our recent study of the mass profiles of seven early-type
galaxies with \chandra\ indicates $c-M$ values consistent with \lcdm\
\citep{hump06b}. The $c-M$ relations for clusters ($M>10^{14}\msun$)
obtained using \xmm\ \citep{poin05a} and \chandra\ \citep{vikh06a}
each are consistent with no variation in $c$ and with the gentle
decline with increasing $M$ expected for CDM models. In each case the
scatter in $\logten\ c$ about the mean relation is not well
constrained and is consistent with the CDM prediction.

Our present investigation aims to improve significantly the
constraints on the $c-M$ relation by analyzing a wider mass range with
many more systems than the previous \chandra\ and \xmm\ studies.  For
this purpose we undertook a program to obtain accurate mass
constraints on relaxed systems with $10^{12}\la M \la 10^{14}\msun$,
representing 24 individual early-type galaxies and galaxy
groups/clusters, each of which possesses high-quality \chandra\ and/or
\xmm\ data \citep{hump06b,gast06a,zapp06a}.  To these systems we added
results for 15 relaxed, massive clusters from \citet{poin05a} and
\citet{vikh06a}. Using this combined sample, optimized for X-ray mass
measurements, we obtain empirical constraints on the local $c-M$
relation with a simple power-law model and compare this model to the
predictions of a suite of CDM models.

We note that the primary purpose of this paper is to assemble accurate
$c-M$ measurements from \chandra\ and \xmm\ observations from galaxy
to cluster scales and to test whether $c$ decreases with increasing
$M$ similar to that predicted by CDM models. However, we also make an
initial attempt to constrain the parameters of CDM models to examine
the efficacy of the $c-M$ relation in this regard and to highlight
sources of systematic error both in the models and data that need to
be addressed with future studies.

The paper is organized as follows. We describe the X-ray $c$ and $M$
measurements in \S \ref{obs}. Our procedure for fitting a power-law
model to the $c-M$ data is discussed in \S \ref{method}. We define the
cosmological models that are compared to the X-ray data in \S
\ref{models}. We present our results in \S \ref{results} and discuss
sources of systematic error in \S \ref{sys}. In \S \ref{conc} we give
our conclusions.  All distance-related quantities, unless stated
otherwise, are computed assuming a flat universe with $\omegam=0.3$,
$\omegalambda=0.7$, and $H_0=100h$~\kmsmpc\ with $h=0.7$.

\section{Observations}
\label{obs}
\begin{table*}[t] \scriptsize
\begin{center}
\caption{Halo Concentration and Virial Mass}
\label{tab.data}
\begin{tabular}{lcc|ccc|ccc|ccc}  \tableline\tableline\\[-7pt]
& & & & & & $M$ & $\sigma_M$ & & & \\
Name & $z$ & $\Delta$ & $c_0$ & $\sigma_{c_0}$ & $\sigma_{\log c_0}$ & ($\msun$)
& ($\msun$) & $\sigma_{\log M}$ & Corr & Corr$_{\rm log}$ & Ref.\\
(1) & (2) & (3) & (4) & (5) & (6) & (7) & (8) & (9) & (10) & (11) & (12)\\
\tableline \\[-7pt]
        NGC4125 &    0.0045 &   101.5 &    10.19 &      1.78 &   7.52e-02 &   6.24e+12 &   8.04e+11 &   6.24e-02 &  -7.87e-01 &  -7.61e-01 &     a\\ 
         NGC720 &    0.0058 &   101.6 &    18.54 &      7.25 &   1.65e-01 &   6.65e+12 &   1.42e+12 &   9.63e-02 &  -7.88e-01 &  -7.83e-01 &     a\\ 
        NGC6482 &    0.0131 &   102.3 &    18.40 &      5.39 &   1.35e-01 &   7.15e+12 &   1.72e+12 &   8.63e-02 &  -8.46e-01 &  -8.87e-01 &     a\\ 
        NGC5129 &    0.0230 &   103.1 &    14.95 &      2.46 &   7.71e-02 &   1.54e+13 &   2.18e+12 &   5.96e-02 &  -8.99e-01 &  -8.94e-01 &     b\\ 
        NGC1407 &    0.0059 &   101.7 &    17.88 &      4.93 &   1.09e-01 &   1.59e+13 &   3.12e+12 &   9.35e-02 &  -8.41e-01 &  -8.35e-01 &     a\\ 
         NGC533 &    0.0185 &   102.7 &    17.32 &      1.21 &   3.14e-02 &   2.29e+13 &   1.11e+12 &   2.05e-02 &  -9.03e-01 &  -9.12e-01 &     b\\ 
        NGC2563 &    0.0149 &   102.4 &    10.16 &      3.43 &   1.16e-01 &   2.61e+13 &   5.46e+12 &   9.78e-02 &  -9.01e-01 &  -9.74e-01 &     b\\ 
          RGH80 &    0.0379 &   104.4 &    10.28 &      1.01 &   4.15e-02 &   2.81e+13 &   1.60e+12 &   2.55e-02 &  -9.23e-01 &  -9.28e-01 &     b\\ 
        NGC1550 &    0.0124 &   102.2 &    17.16 &      1.12 &   2.80e-02 &   3.12e+13 &   1.49e+12 &   2.08e-02 &  -9.53e-01 &  -9.58e-01 &     b\\ 
        NGC4472 &    0.0033 &   101.4 &    13.07 &      1.64 &   5.09e-02 &   3.27e+13 &   3.99e+12 &   5.97e-02 &  -8.90e-01 &  -8.89e-01 &     a\\ 
        NGC4325 &    0.0257 &   103.3 &    11.46 &      1.28 &   5.61e-02 &   3.50e+13 &   8.10e+12 &   8.34e-02 &  -9.71e-01 &  -9.83e-01 &     b\\ 
        NGC4649 &    0.0037 &   101.5 &    20.80 &      2.45 &   5.17e-02 &   3.50e+13 &   4.62e+12 &   6.29e-02 &  -8.88e-01 &  -8.63e-01 &     a\\ 
        NGC5044 &    0.0090 &   101.9 &    11.32 &      0.29 &   1.09e-02 &   3.73e+13 &   1.17e+12 &   1.40e-02 &  -9.50e-01 &  -9.53e-01 &     b\\ 
         IC1860 &    0.0223 &   103.1 &     9.59 &      0.88 &   3.91e-02 &   5.10e+13 &   4.62e+12 &   3.98e-02 &  -9.66e-01 &  -9.72e-01 &     b\\ 
           MKW4 &    0.0200 &   102.9 &    12.72 &      0.86 &   2.96e-02 &   6.22e+13 &   3.38e+12 &   2.39e-02 &  -9.43e-01 &  -9.46e-01 &     b\\ 
        NGC4261 &    0.0075 &   101.8 &     3.72 &      0.86 &   1.09e-01 &   6.69e+13 &   1.03e+13 &   6.35e-02 &  -8.29e-01 &  -7.95e-01 &     a\\ 
 RXJ1159.8+5531 &    0.0810 &   108.0 &    11.48 &      2.86 &   1.28e-01 &   9.10e+13 &   6.89e+13 &   1.73e-01 &  -8.12e-01 &  -9.79e-01 &     b\\ 
           A262 &    0.0163 &   102.5 &     8.87 &      0.67 &   3.46e-02 &   1.11e+14 &   1.06e+13 &   3.99e-02 &  -9.77e-01 &  -9.83e-01 &     b\\ 
  MS0116.3-0115 &    0.0452 &   105.0 &     6.57 &      2.21 &   1.41e-01 &   1.28e+14 &   8.04e+13 &   2.14e-01 &  -8.35e-01 &  -9.72e-01 &     b\\ 
     ESO5520200 &    0.0314 &   103.8 &     7.79 &      0.75 &   4.46e-02 &   1.31e+14 &   1.96e+13 &   6.06e-02 &  -9.53e-01 &  -9.57e-01 &     b\\ 
           MKW9 &    0.0498 &   105.4 &     7.48 &      0.93 &   5.41e-02 &   1.44e+14 &   3.61e+13 &   1.11e-01 &  -6.49e-01 &  -9.36e-01 &     d\\ 
           AWM4 &    0.0317 &   103.9 &     9.27 &      0.84 &   3.72e-02 &   1.62e+14 &   1.82e+13 &   5.10e-02 &  -9.18e-01 &  -9.33e-01 &     b\\ 
     ESO3060170 &    0.0358 &   104.2 &     9.17 &      1.04 &   6.12e-02 &   1.81e+14 &   7.56e+13 &   1.09e-01 &  -8.84e-01 &  -9.76e-01 &     b\\ 
          A2717 &    0.0490 &   105.3 &     6.45 &      0.36 &   2.46e-02 &   1.83e+14 &   1.20e+13 &   2.84e-02 &  -9.79e-01 &  -9.83e-01 &     b\\ 
          A1991 &    0.0592 &   106.2 &     8.91 &      0.66 &   3.22e-02 &   1.93e+14 &   2.67e+13 &   6.04e-02 &  -1.00e+00 &  -1.00e+00 &     e\\ 
          A1983 &    0.0442 &   104.9 &     5.34 &      0.99 &   8.15e-02 &   1.99e+14 &   7.62e+13 &   1.76e-01 &  -2.87e-01 &  -3.92e-01 &     d\\ 
          A2589 &    0.0414 &   104.7 &     6.36 &      0.32 &   2.33e-02 &   3.24e+14 &   2.37e+13 &   3.00e-02 &  -9.56e-01 &  -9.66e-01 &     c\\ 
          A2597 &    0.0852 &   108.4 &     8.26 &      0.70 &   3.71e-02 &   3.56e+14 &   3.91e+13 &   4.80e-02 &  -7.87e-01 &  -1.00e+00 &     d\\ 
           A383 &    0.1883 &   116.8 &     9.61 &      0.89 &   4.03e-02 &   4.70e+14 &   4.76e+13 &   4.41e-02 &  -5.12e-01 &  -1.00e+00 &     e\\ 
           A133 &    0.0569 &   106.0 &     6.72 &      0.61 &   3.97e-02 &   5.37e+14 &   6.44e+13 &   5.23e-02 &  -5.50e-01 &  -1.00e+00 &     e\\ 
          A1068 &    0.1375 &   112.7 &     5.44 &      0.38 &   3.07e-02 &   6.96e+14 &   6.01e+13 &   3.76e-02 &  -9.42e-01 &  -1.00e+00 &     d\\ 
           A907 &    0.1603 &   114.5 &     7.75 &      0.94 &   5.27e-02 &   7.39e+14 &   5.99e+13 &   3.53e-02 &  -3.87e-01 &  -1.00e+00 &     e\\ 
          A1795 &    0.0622 &   106.5 &     6.80 &      0.38 &   2.44e-02 &   1.02e+15 &   8.78e+13 &   3.75e-02 &  -6.48e-01 &  -1.00e+00 &     e\\ 
        PKS0745 &    0.1028 &   109.9 &     7.32 &      0.57 &   3.40e-02 &   1.20e+15 &   1.44e+14 &   5.24e-02 &  -2.64e-01 &  -1.00e+00 &     d\\ 
           A478 &    0.0881 &   108.6 &     7.61 &      0.58 &   3.29e-02 &   1.25e+15 &   1.65e+14 &   5.74e-02 &  -2.28e-01 &  -1.00e+00 &     e\\ 
          A2029 &    0.0779 &   107.8 &     8.47 &      0.44 &   2.26e-02 &   1.27e+15 &   1.18e+14 &   4.02e-02 &  -4.19e-01 &  -1.00e+00 &     e\\ 
          A1413 &    0.1429 &   113.1 &     6.56 &      0.38 &   2.52e-02 &   1.29e+15 &   1.29e+14 &   4.37e-02 &  -4.41e-01 &  -1.00e+00 &     e\\ 
          A2204 &    0.1523 &   113.9 &     6.77 &      0.55 &   3.51e-02 &   1.41e+15 &   1.55e+14 &   4.80e-02 &  -2.55e-01 &  -1.00e+00 &     d\\ 
          A2390 &    0.2302 &   120.0 &     4.13 &      0.32 &   3.41e-02 &   2.13e+15 &   2.14e+14 &   4.38e-02 &  -3.13e-01 &  -1.00e+00 &     e\\ 
\tableline \\
\end{tabular}
\tablecomments{Col.(1): Cluster name. Col.(2) Redshift. Col.(3)
Reference overdensity for virial radius and mass definition. Col.(4)
Scaled concentration $c_0 = (1+z)c$. Col.(5) Standard deviation
on $c_0$. Col.(6) Standard deviation on $\log_{10} c_0$. Col.(7)
Virial mass. Col.(8) Standard deviation on $M$. Col.(9) Standard
deviation on $\log_{10} M$. Col.(10) Correlation coefficient for $c_0$
and $M$. Col.(11) Correlation coefficient for $\log_{10} c_0$ and
$\log_{10} M$. Col.(12) Reference for $c$ and $M$ values: a --
\citet{hump06b}, b --
\citet{gast06a}, c -- \citet{zapp06a}, d --
\citet{poin05a}, e -- \citet{vikh06a}. Note that literature results
obtained at different $\Delta$ values have been converted to those in
column (3) as explained in \S \ref{obs}.}
\end{center}
\end{table*}
The measurements of concentrations and virial masses of 23 early-type
galaxies and galaxy groups \citep{hump06b,gast06a} form the core of
our sample. In these papers we considered only those systems with the
highest quality \chandra\ and/or \xmm\ data. Moreover, to insure that
hydrostatic equilibrium is a good approximation, we visually inspected
all the early-type galaxies and groups in the public data archives and
selected those systems that possessed the most regularly shaped X-ray
images devoid of strong asymmetries. 

To these we added several clusters to populate the high-mass portion
of the $c-M$ diagram. Firstly, we include our analysis of the
radio-quiet (and very symmetrical) cluster A2589
\citep{zapp06a}. Secondly, we added clusters from the \xmm\
study by \citet{poin05a} and the \chandra\ study by \citet{vikh06a},
each of whom also focused on the highest quality observations of the
most relaxed systems. For clusters that are common to both studies, we
used results from \citet{vikh06a} because of more accurate background
subtraction and temperature constraints, the latter resulting from the
much smaller point spread function of \chandra; i.e., less biased
temperature measurements in the presence of strong temperature
gradients. For those few (lowest mass) systems in the \citet{poin05a}
and \citet{vikh06a} studies that overlap with our work cited above, we
use our values. 

The final sample (see Table \ref{tab.data}) consists of 39 systems
spanning the mass range $(0.06-20)\times 10^{14}\msun$, with $c$ and
$M$ values inferred using the NFW dark matter profile. We follow the
convention that defines the virial radius (\rvir) so that the mean
density within \rvir\ equals $\Delta\rho_c$, where $\rho_c(z)$ is the
critical density of the universe at redshift $z$, and $\Delta$ is
obtained from the solution to the top-hat spherical collapse model.
To evaluate $\Delta$ we use the approximation obtained by
\citet{brya98a}.  For simplicity, when considering dark energy (DE)
models with $w \ne -1$, we adopt the value of $\Delta$ appropriate for
$w = -1$.  Although technically the virial overdensity changes with $w
\ne -1$ \citep{kuhl05a}, this choice has no effect on our conclusions.  The
values of $\Delta$ for the concordance cosmology at the appropriate
redshift for each system are listed in Table \ref{tab.data}.

For results obtained using $\Delta$ values different from those in
Table \ref{tab.data}, we converted $c$ and $M$ to our adopted $\Delta$
using the formula of \citet{hukr03a}, which assumes an NFW mass
profile. We converted the standard deviations for $c$ and $M$ in one
of two ways. For those systems taken from our previous work, we have
at least 20 Monte Carlo error simulations. Hence, we convert each
simulated value of $c$ and $M$ to the new $\Delta$ and then compute
the standard deviation of the converted values. This procedure
provides a self-consistent conversion of the standard deviations.  For
systems taken from the literature, we simply convert the lower limit
(e.g., $c - \sigma_c$) and upper limit (e.g., $c + \sigma_c$) each to
the new $\Delta$. The new standard deviation is then set to one-half
the difference of the converted upper and lower limits.  (By
performing this procedure on our own data and comparing to the
rigorous method using the error simulations, we find that this
procedure is adequate for our present study.)

Since we perform our analysis of the $c-M$ relation in log space (see
below), we also require the standard deviations of $\log_{10} c$ and
$\log_{10} M$. As above, for systems analyzed in our previous studies,
we compute self-consistently the log errors using Monte Carlo error
simulations. The other systems use the simple prescription noted above,
but using the log values. The log errors on $c$ and $M$ are listed in Table
\ref{tab.data} as, respectively, $\sigma_{\log c}$ and $\sigma_{\log
M}$. 

For the systems with Monte Carlo error simulations we also computed
the covariance between $c$ and $M$ (and $\log_{10} c$ and $\log_{10}
M$) for the different systems; e.g., $\sum (c_i
-\langle c\rangle)(M_i - \langle M\rangle)$, where the brackets
represent the mean quantity, and the sum is over the simulations. In Table
\ref{tab.data} we report these as the correlation coefficient, which
is the covariance divided by $\sigma_c\sigma_M$ (or $\sigma_{\log
c}\sigma_{\log M}$). For the objects from the literature we used the
average covariance obtained from the others to evaluate the
correlation coefficient, setting the coefficient to -1 if a value
$<-1$ was obtained. Most of the correlation coefficients have values
near -1 indicating that $c$ and $M$ are anti-correlated. 

Finally, the results listed in Table \ref{tab.data} for the
concentration parameter actually refer to the quantity, $c_0=(1+z)c$,
since this quantity is what is fitted in our analysis below. Note that
$c_0$ is also used when computing the covariance and correlation
coefficient.

\section{Analysis Method}
\label{method}

We focus our analysis on a simple power-law representation of the
$c-M$ data,
\begin{equation}
c = \frac{c_{14}}{1+z}\left(\frac{M}{M_{14}}\right)^{\alpha},
\label{eqn.cm} 
\end{equation}
where $z$ is the redshift, and both $c_{14}$ and $\alpha$ are
constants independent of $M$. We set the reference mass to,
$M_{14}=10^{14}h^{-1}\msun$, which lies close to the midpoint (in log
space) of the mass range of our sample. CDM models generally predict
that $c$ decreases with increasing $M$ (e.g., B01; D04;
\citealt{kuhl05a,macc06a}).  A key goal of our study, therefore, is to
determine whether $\alpha<0$.  For the most massive systems ($M\ga
3\times 10^{14}\msun$) in CDM models D04 find that $c_{14}$ is much
more sensitive to cosmological parameter variations than $\alpha$ (see
\S \ref{models}).

To constrain the parameters of eqn.\ [\ref{eqn.cm}] it is necessary to
account for the error estimates on both $c$ and $M$. Consequently, we
employ the BCES method (i.e., bivariate correlated errors with
intrinsic scatter) described by \citet{akri96} to estimate $\alpha$,
$c_{14}$, and the intrinsic scatter about the best relation.  We
performed the BCES fitting using software kindly provided by
M. Bershady\footnote{http://www.astro.wisc.edu/$\sim$mab/archive/stats/stats.html}.
Since BCES is a linear regression method we transform eqn.\
[\ref{eqn.cm}] to the form, $y = \alpha x + b$, where $x \equiv
\log_{10} M$, $y\equiv \log_{10} c_0 = \log_{10} (1+z)c$, and $b\equiv
\log_{10}(c_{14}/M_{14}^{\alpha})$. The parameter $c_{14}$ is derived
from $\alpha$ and $b$: $c_{14}=10^{b}M_{14}^{\alpha}$.  Since the
fractional error $\sigma_{\log M}/\log_{10}M$ is typically smaller by
more than an order of magnitude of the fractional error, $\sigma_{\log
c_{0}}/\log_{10} c_{0}$ (see Table \ref{tab.data}), we always use
$x=\log_{10}M$ as the independent variable. By default we also use the
Corr$_{\rm log}$ values in Table \ref{tab.data} for the BCES
method. Despite the strong anti-correlation between $c$ and $M$, we
find that including the Corr$_{\rm log}$ values in the analysis has an
insignificant impact on the estimated $\alpha$ and $c_{14}$ values.

We determine the best estimate of $\alpha$ and $c_{14}$ by performing the
BCES method on $10^6$ bootstrap resamplings of the data. We take the
mean of the bootstrap simulations to be the best estimates and
construct error ellipses about these best values. The confidence
contour spacings computed from the bootstrap simulations correspond
quite closely to the $\Delta\chi^2$ values for two parameters assuming
normally distributed errors.

We estimate the intrinsic scatter on the concentration as,
\begin{eqnarray}
\left(\sigma_y^2\right)_{\rm int} & = & \left(\sigma_y^2\right)_{\rm
total} - \left(\sigma_y^2\right)_{\rm stat} \nonumber\\ 
& = & \frac{1}{N}\sum_{i=1}^N\left(y_i - y_i^{model}\right)^2 -
\frac{1}{N}\sum_{i=1}^N \sigma_{y_i}^2, \label{eqn.scat}
\end{eqnarray}
where $N$ is the number of data points
$(x_i,y_i)=(\log_{10}M_i,\log_{10}{c_{0,i}})$ corresponding to the
entries in Table \ref{tab.data}; $y_i^{model} = \alpha x_i + b$, where
$\alpha$ and $b$ are the best (mean) estimates from the bootstrap
simulations; and $\sigma_{y_i} = \sigma_{\log {c_0}}$ in Table
\ref{tab.data}.

\section{Cosmological Models}
\label{models}
\begin{table}[t] \footnotesize
\begin{center}
\caption{Cosmological Model Parameters}
\label{tab.models}
\begin{tabular}{lccccccc}  \tableline\tableline\\[-7pt]
Name & \omegam\ & \omegalambda\ & $\Omega_Bh^2$ & $h$ & $\sigma_8$ &
$n_s$ & -w\\ \tableline \\[-7pt]
\lone\   & 0.30 & 0.70 & 0.022 & 0.7 & 0.90 & 1.00 & 1.0       \\
\lthree\ & 0.24 & 0.76 & 0.022 & 0.73 & 0.76 & 0.96 & 1.0       \\
DECDM    & 0.30 & 0.70 & 0.022 & 0.7 & 0.90 & 1.00 & 0.6     \\
QCDM     & 0.30 & 0.70 & 0.022 & 0.7 & 0.82 & 1.00 & $\approx 0.8$ \\
OCDM     & 0.30 & 0.00 & 0.022 & 0.7 & 0.90 & 1.00 & $\cdots$ \\
\tableline \\
\end{tabular}
\tablecomments{\omegam\ is the energy density parameter for matter in
the universe; \omegalambda\ is the energy density parameter associated
with a cosmological constant or, more generally, dark energy;
$\Omega_B$ is the energy density parameter of baryons; $h$ is
$H_0$/100~\kmsmpc; $\sigma_8$ is the rms mass fluctuation within
spheres of comoving radius $8h^{-1}$~Mpc. See \S \ref{models}.}
\end{center}
\end{table}
In Table \ref{tab.models} we define the CDM-based cosmological models
to be compared with the observations. \lone\ is the standard
concordance model, which essentially reflects the combined constraints
from the first year of WMAP CMB observations, the supernovae Hubble
diagram, and the large-scale clustering of galaxies
\citep[e.g.,][]{sper03a}. \lthree\ effectively updates \lone\ using
the third year of WMAP \citep{sper06a}. Each of these models assumes a
constant dark energy equation of state with $w=-1.0$ (i.e., a
cosmological constant).  In addition, we consider a dark energy model
with $w=-0.6$ (DECDM) and a quintessence model (QCDM) with a
Ratra-Peebles potential \citep[and references
therein]{ratr88a,peeb03a} as implemented in D04. The QCDM model has a
nearly constant $w(z)\approx -0.8$ (see Fig.\ 1 of D04) and a lower
\sige.  Finally, we include an open model (OCDM) with the same
parameters as \lone\ except with $\omegalambda=0$.

The median relation between concentration and virial mass as a
function of redshift for CDM halos is described well by the
semi-analytic model proposed by B01,
\begin{equation}
c(M,z) = K\frac{1+z_c}{1+z}, \label{eqn.cm.b01}
\end{equation}
where $K$ is the normalization constant, $z$ is the redshift of the
halo, and $z_c$ is the redshift when the halo ``collapsed''. This
collapse redshift is defined implicitly by the equation,
\begin{equation}
\sigma(FM) = \delta_c(z_c),
\end{equation}
where $\delta_c$ is the equivalent linear overdensity for spherical
collapse at $z_c$, $\sigma$ is the $z=0$ linear rms density fluctuation
corresponding to a mass $FM$, and $F$ is a constant. (Note that eqn.\
[\ref{eqn.cm.b01}] is not a power-law, though over small mass ranges
-- within a factor of 5-10 -- the B01 model can be well approximated
by one.) The constants $F$ and $K$ must be specified by comparison
with numerical simulations.

For halos with $M\la 10^{13}h^{-1}\msun$ it is found that $F=0.01$ and
$K\approx 3.5$ provide the best description of the results from N-body
simulations, although $F=0.001$ and $K \approx 3$ is acceptable (B01;
\citealt{kuhl05a,macc06a}).  Larger box simulations that focus on
higher mass ranges clearly prefer $F=0.001$.  One of us (J. Bullock)
finds that $F=0.001$ and $K=3.1$ matches the $c-M$ results for CDM
halos ($0.6-2.5 \, \times 10^{14}h^{-1}\msun$) simulated by
\citet{tasi04a}.  For simulations of even more massive clusters
($3.1-17 \, \times 10^{14}h^{-1}\msun$), D04 find higher
concentrations and that $K=3.5$ is required for the same
$F$. Consequently, when comparing CDM models to the full mass range of
the X-ray data in Table
\ref{tab.data} we shall consider the $c-M$ relations obtained with the
B01 model using $F=0.001$ with both $K=3.1$ and $K=3.5$. (In \S
\ref{fk} we discuss further the different $F$ and $K$ values obtained
by different investigators.)
\begin{table}[t] \footnotesize
\begin{center}
\caption{Power-Law Approximation for CDM Clusters}
\label{tab.models.cm}
\begin{tabular}{lcc}  \tableline\tableline\\[-7pt]
Name & $\alpha$ & $c_{14}$\\ \tableline \\[-7pt]
\lone\   & -0.104 & 7.63\\
DECDM    & -0.094 & 9.05\\
QCDM     & -0.111 & 7.39\\
OCDM     & -0.091 & 11.47\\
\lone$\rm _B$   & -0.142 & 8.43\\
\lthree$\rm _B$ & -0.155 & 6.23\\
\tableline \\
\end{tabular}
\tablecomments{Power-law parameters (eqn.\ \ref{eqn.cm}) for the theoretical
relation between concentration and virial mass of high-mass clusters
obtained by D04 for the \lone, DECDM, QCDM, and OCDM models. We
converted the results presented in Table 2 of D04 to our definition of
the virial radius at $z=0$ (i.e., $\Delta=101.1$). The
\lone$\rm _B$ and \lthree$\rm _B$ models are the same as the \lone\
and \lthree\ models, but we computed the power-law parameters using
the B01 model with $F=0.001$ and $K=3.5$. See \S \ref{models}.}
\end{center}
\end{table}
In the high-mass ``cluster'' regime D04 find that the $c-M$ relation
for CDM models is adequately parameterized by the simple power-law
model given by eqn.\ [\ref{eqn.cm}]. We list the power-law parameters
obtained by D04 in Table \ref{tab.models.cm} but converted to our
definition of virial radius. (D04 use $\Delta=200\rho_b=60\rho_c$ for
$\omegam=0.3$.)  That is, initially we compute $c(M)$ using
$c_{14}$ and $\alpha$ from D04 (at $z=0$). Then we convert the resulting $c$ and
$M$ values to $\Delta=101.1$ using the approximation of
\citet{hukr03a}. These values are used to compute the slope and
$c_{14}$, which are listed in Table \ref{tab.models.cm}.  The
parameter $c_{14}$ decreases by $\approx 20\%$ while $\alpha$ remains
nearly constant under this transformation.  For comparison, in Table
\ref{tab.models.cm} we also list $\alpha$ and $c_{14}$ computed from
the tangent line to the $c-M$ profiles located at $M=M_{14}$ of the
\lone\ and \lthree\ models obtained using the B01 approach with
$F=0.001$ and $K=3.5$. These results are listed as
\lone$\rm _B$ and \lthree$\rm _B$ respectively. 
(We note that the virial quantities for the X-ray data in Table
\ref{tab.data} all refer to the \lone\ cosmology. For consistent
comparison, we also use the \lone\ cosmology when converting all the
D04 models mentioned above.)

The concentration is sensitive to the formation time and dynamical
state of the halo (e.g., \citealt{nfw,eke01a,jing00a}; B01;
\citealt{wech02a,macc06a}).   For  example, \citet{jing00a} and
\citet{macc06a} find that their ``relaxed''   dark halo samples  have
concentrations that are systematically larger by $\sim 10\%$ compared
to the whole population.  \citet{wech02a} found that the halo
concentration at fixed mass is set almost exclusively by the
``formation epoch'' of the halo (the time when the mass accretion rate
of the halos is slowed below a critical value).  Indeed, when
\citet{wech02a} focused on halos with no major mergers since $z = 2$
they found $\sim 10 \%$ higher concentrations for that population.
Therefore, a proper comparison between the theoretical $c-M$ relation
with observations requires the observed and simulated halos be
selected in a consistent manner. The X-ray data presented in Table
\ref{tab.data} represent objects that were selected to be the most
relaxed, X-ray--bright, systems.  But the theoretical $c-M$ models
discussed above (Table \ref{tab.models.cm}) were obtained using all
available halos in the N-body simulations. Consequently, we consider
this source of systematic error in our comparisons of the CDM models
with the X-ray data.

Finally, the magnitude of the intrinsic scatter about the median
theoretical $c-M$ relation does not vary over a large class of CDM
models and is independent of $M$. The value of $\sigma_{\log c}\approx
0.14$ ($\log\equiv\logten$) obtained by B01 has been found by several
independent investigations (\citealt{jing00a,tasi04a}; D04;
\citealt{macc06a}); note this consistency is observed after accounting
for the fact that $\ln$, rather than \logten, is used in these other
studies. Early forming, relaxed, halos tend to exhibit smaller scatter
$\sigma_{\log c}\approx 0.10$ \citep{jing00a,wech02a,macc06a}.

\vskip 1cm

\section{Results}
\label{results}
\begin{figure*}[t]
\parbox{0.49\textwidth}{
\centerline{\includegraphics[height=0.35\textheight]{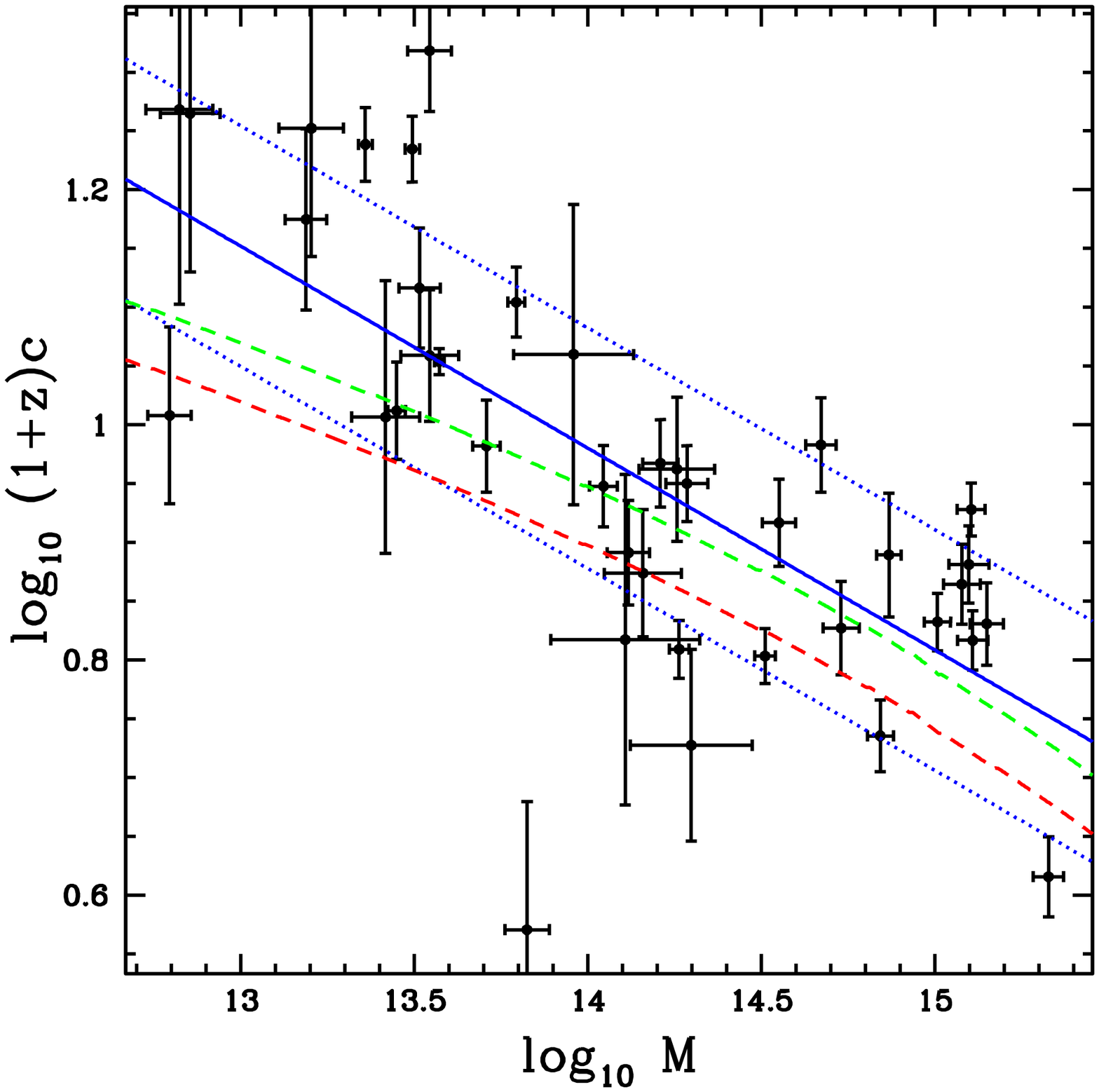}}}
\parbox{0.49\textwidth}{
\centerline{\includegraphics[height=0.35\textheight]{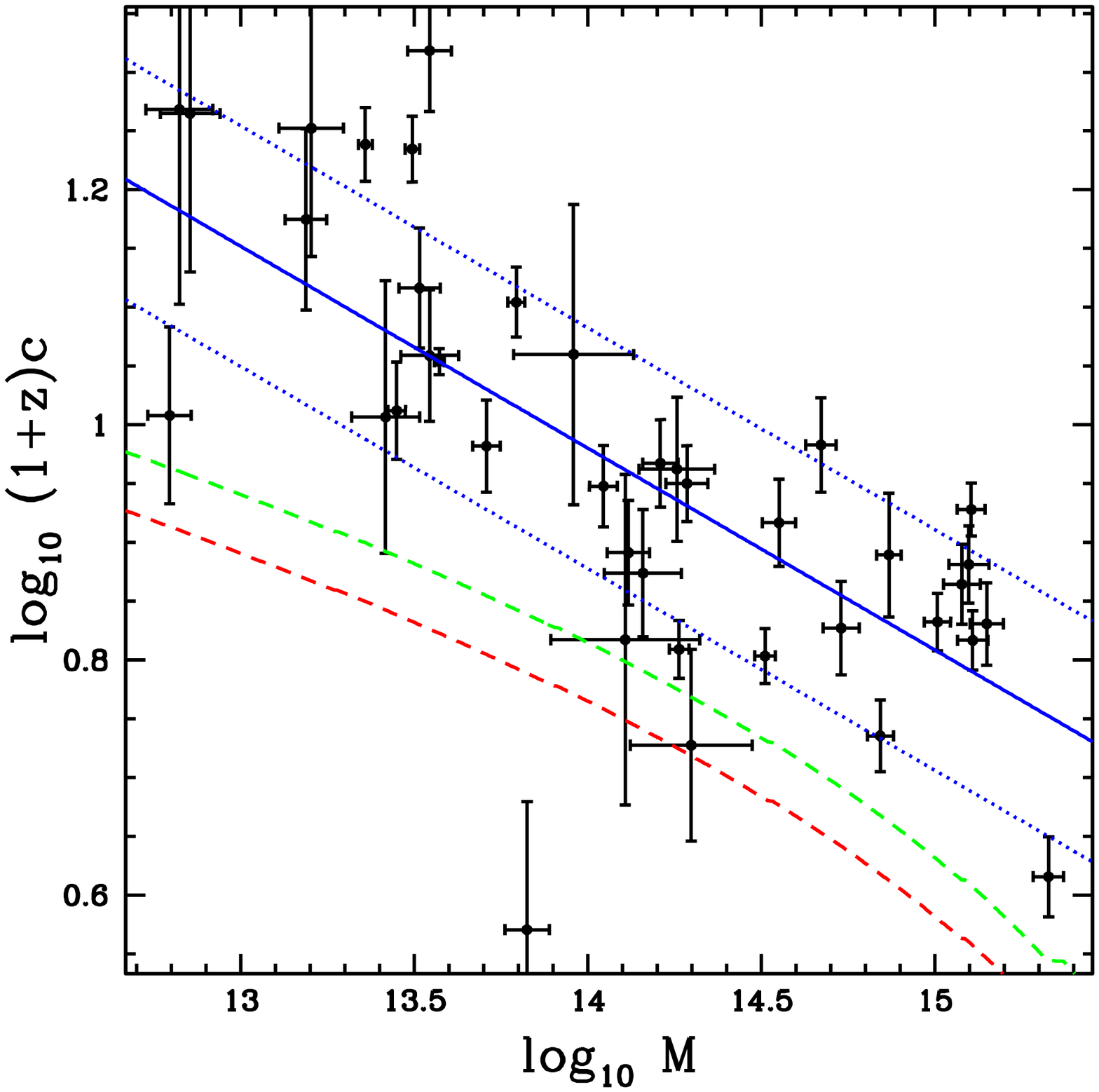}}}
\caption{\label{fig.cm} 
Results of fitting eqn.\ \ref{eqn.cm} in log space to the entire
sample. Displayed are the best-fitting model (solid blue line), the
$1\sigma$ intrinsic scatter (dotted blue lines), and the predicted
relation obtained from cosmological simulations for the ({\sl Left
panel}) \lone\ and ({\sl Right panel}) \lthree\ models (dashed
lines). The two dashed lines attempt to represent fits to different
mass ranges in the cosmological simulations as explained in the text
(\S \ref{models}). The lower (red) dashed line refers to fits of halos
up to $M\sim 0.3\times 10^{15}\msun$ ($K=3.1$) while the upper
(green) dashed line refers to fits of higher mass halos $M\approx
(0.3-1)\times 10^{15}\msun$ ($K=3.5$) obtained by D04.}
\end{figure*}

\begin{figure*}[t]
\parbox{0.49\textwidth}{
\centerline{\includegraphics[height=0.35\textheight]{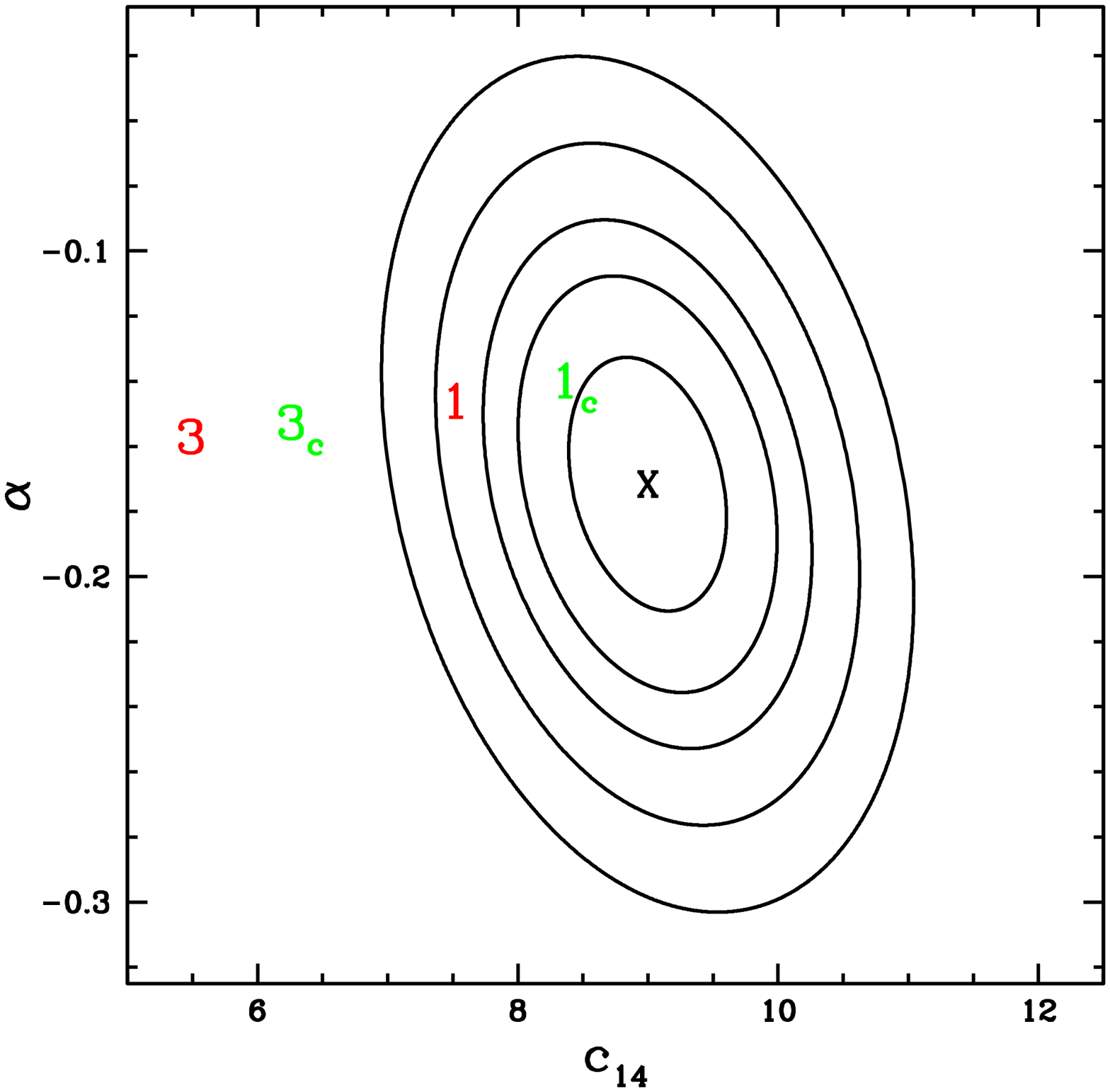}}}
\parbox{0.49\textwidth}{
\centerline{\includegraphics[height=0.35\textheight]{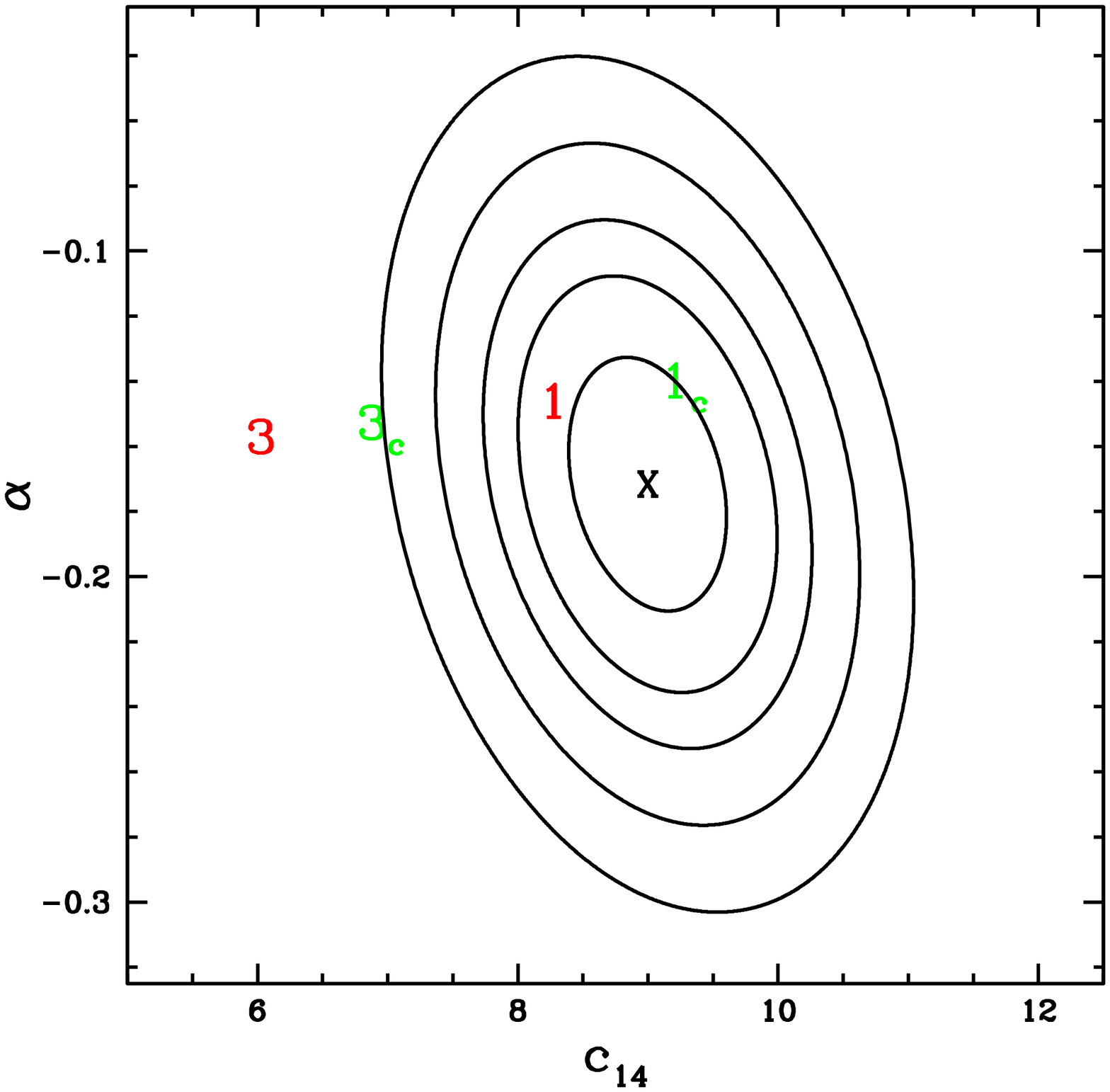}}}
\caption{\label{fig.errors} 
Best-fitting value (X) and confidence contours (68\%, 95\%, 99\%,
99.9\%, 99.99\%) for $\alpha$ and $c_{14}$ obtained from the entire
sample. ({\sl Left panel}) As in Fig.\ \ref{fig.cm} results for the
\lone\ (green 1) and \lthree\ (red 3) models are displayed. Models
with subscript ``c'' refer to the higher normalization $K=3.5$ for the
cluster regime preferred by D04. ({\sl Right panel}) Models have
$c_{14}$ increased by 10\% to represent relaxed, early-forming halos
(see \S \ref{models} and \citealt{macc06a}).}
\end{figure*}

In \S \ref{sys} we discuss possible sources of systematic error both
in the measurements from the X-ray data and the theoretical
predictions. Possible systematic errors affecting the X-ray mass
measurements of our sample of relaxed galaxies and clusters are
expected to be small and do not change the conclusions of the analysis
presented here (see beginning of \S \ref{sys} and \S \ref{he}).

\subsection{All Halos}
\label{all}

When fitting the power-law relation (eqn.\ \ref{eqn.cm}) to the entire
sample we obtain $\alpha=-0.172\pm 0.026$ and $c_{14}=9.0\pm 0.4$
(quoted errors $1\sigma$). The estimated intrinsic scatter in
$\logten\ (1+z)c$ is $0.102\pm 0.004$ (eqn.\ \ref{eqn.scat}), with a
total scatter of 0.12. In Figure \ref{fig.cm} we plot the best-fitting
model with intrinsic scatter. Inspection of Figure \ref{fig.cm}
reveals that the power-law with intrinsic scatter is a good
approximate representation of the X-ray results, especially for
$\logten\ M > 13.5$. For lower masses the X-ray measurements tend to
lie above the best-fitting model, though most are consistent with
lying within the $1\sigma$ range covered by the intrinsic scatter.
Using equation [3] of \citet{trem02a} we compute a reduced $\chi^2$
value of 1.07 (37 dof) for the fit, confirming the visual impression
that the power law is a good, but not perfect, representation of the
X-ray data.

The slope $\alpha$ is constrained to be negative at the $6.6\sigma$
level, demonstrating at high significance that $c$ decreases with
increasing $M$, as expected in CDM models. In Figure \ref{fig.cm} we
also plot the theoretical $c-M$ relation for the \lone\ model for both
$K=3.1$ and $K=3.5$ as discussed in \S \ref{models}.  For $\logten\ M
> 14$ the $K=3.5$ model recommended by D04 for clusters lies very
close to, but just below, the empirical power-law fit.  Allowing for a
systematic $\approx 10\%$ increase in concentration for relaxed, early
forming halos \citep{jing00a,wech02a,macc06a} lifts the $K=3.5$ model
slightly over the power-law, with a marginal improvement in the
agreement between the models. A recent theoretical study
\citep{wang06a} finds that the higher mass halos experience less of
bias associated with relaxation and formation time.  In fact,
systematically increasing the concentrations of the $K=3.5$ model by
$\approx 7\%$ would provide a closer match to the empirical power-law
fit to the X-ray data.

For $\logten\ M <14$ the $K=3.5$ model must quickly transform into the
$K=3.1$ model as $M$ decreases from the cluster to the galaxy/group
regime as discussed in \S \ref{models}. But in this mass range the
$K=3.1$ model lies consistently below the power-law model and is
similar to the $1\sigma$ lower limit on the power-law given by the
intrinsic scatter. Increasing the concentrations of the $K=3.1$ model
by $\approx 10\%$ for relaxed halos improves the agreement with the
power-law, and has the effect of essentially pushing the $K=3.1$
model into the $K=3.5$ model.  With this accounting for the systematic
bias arising from relaxed, early forming halos the \lone\ model is an
equally good representation of the X-ray data for masses $\logten\ M >
13.5$. For the lowest masses, the
\lone\ model still lies below the power-law (within the $1\sigma$
intrinsic scatter) and may represent a real discrepancy.

We summarize this qualitative discussion with a quantitative
comparison of the \lone\ model with the data as represented by the
results of the empirical power-law fit. (We do not fit the B01 model
to the data for the following reasons. First, it is non-trivial to
account for the error bars in both coordinates when fitting a general
model; the Akritas \& Bershady method only applies to linear
regression. Second, the prediction of the B01 model possesses
significant systematic uncertainty in the parameter $K$ (at least for
masses $<10^{14}\msun$) as discussed above. By showing the B01 model
predictions separately for the range of interesting K values, without
fitting, we provide a clear demonstration of the importance of
reducing this systematic uncertainty with future theoretical
studies. When that happens, formal fitting of the B01 model will be
investigated.)  In Figure \ref{fig.errors} we show the error contours
estimated for $\alpha$ and $c_{14}$ from the bootstrap simulations (\S
\ref{method}). We calculated the slope and concentration of the $c-M$
relation of the \lone\ model at $10^{14}h^{-1}\msun$ for both the
$K=3.1$ and $K=3.5$ cases. (The values for $K=3.5$ are listed in Table
\ref{tab.models.cm}.) These results are plotted in Figure
\ref{fig.errors} as ``1'' for $K=3.1$ and $1_c$ for $K=3.5$, the
latter representing the ``cluster'' regime. The right panel of Figure
\ref{fig.errors} increases the concentrations by 10\% to represent
relaxed, early forming systems, and shows the \lone\ model lies near
the $68\%$ confidence contour.  This level of agreement applies for
masses $\logten\ M > 13.5$. As discussed above, the agreement is worse
at the lowest masses. However, performing a similar comparison at
$10^{13}h^{-1}\msun$ we find the local \lone\ slope and normalization
lie just within the 95\% confidence contour of the power-law model
(both with and without the 10\% correction for relaxed systems); i.e.,
the disagreement is not very significant even at the low-mass end. If
real, the discrepancy may signify very early forming fossil groups in
the sample \citep{dong05a}.

The intrinsic scatter of 0.10 in $\logten\ (1+z)c$ obtained for the
power-law fit is smaller than the value of 0.14 for all dark matter
halos in CDM simulations but agrees extremely well with the value
expected for the most relaxed, early forming systems
\citep{jing00a,wech02a,macc06a}.  Consideration of this result for the
scatter and the average $c-M$ relation above, we conclude that the
\lone\ model is consistent with the X-ray data, provided the X-ray
sample reflects the most relaxed, early forming systems in the
population. (This corroborates our selection criteria discussed in \S
\ref{obs}.)

Now we perform the analogous comparison of the \lthree\ model with the
X-ray data and associated power-law fit. The $c-M$ relation is
displayed in the right panel of Figure \ref{fig.cm}. The \lthree\
model lies well-below the power-law at all masses, even when
allowing for the expected 
10\% increase in concentration for relaxed halos.  In
Figure \ref{fig.errors} we compare the \lthree\ model at
$10^{14}h^{-1}\msun$ with the empirical power-law. Even when
considering the 10\% increase in concentrations for relaxed halos, the
\lthree\ model lies on the 99.99\% contour.
Systematic errors associated with the X-ray measurements cannot
explain this level of disagreement (see \S \ref{sys}).

The key parameter responsible for the poor performance of the \lthree\
model with respect to \lone\ is the low value of \sige\ (0.76). In
order to bring the \lthree\ model within the 99\% contour in the right
panel of Figure \ref{fig.errors} requires $\sige>0.84$ where we have
kept the other cosmological parameters fixed to their values in Table
\ref{tab.models}. This limit is conservative since (1) we use
$K=3.5$, (2) we assume a full 10\% upward shift for the bias from
relaxed, early forming systems, which may be less for massive
clusters, and (3) we have approximated the B01 models as power-laws
using their predictions only near $10^{14}h^{-1}\msun$. The
sensitivity of the concentrations to \sige\ results from the impact
that \sige\ has on the average halo formation times
\citep[e.g.,][]{eke01a,alam02a,bosc03a}.

Other cosmological parameters, however, contribute to the large
discrepancy of the \lthree\ model. If we remove the tilt of the power
spectrum (i.e., set $n_s=1$) then we obtain $\sige>0.80$ at the 99\%
confidence level. Finally, if we further set $\omegam=0.3$ and
$h=0.73$ (so the model is the same as \lone\ except with variable
\sige), then the 99\% constraint falls to $\sige>0.76$, the value
associated with the \lthree\ model. (We reiterate this limit is
conservative as noted above. Similarly, we obtain $\sige<1.07$ at 99\%
confidence, where the $K=3.1$ model is used here to be conservative.)
Hence, although the lower value of \sige\ is the primary cause of the
poor performance of the \lthree\ model, the combined action of the
power-spectrum tilt with the lower value of \omegam\ exacerbate the
discrepancy with the X-ray $c-M$ relation.

If the \lthree\ parameters are correct, particularly the low value of
\sige, then a fundamental modification of the model is required to
increase the concentration values to match the X-ray results. Both D04
and \citet{kuhl05a} have shown that changing the dark energy equation
of state parameter ($w$) has the effect of systematically raising
(larger $w$) or lowering (smaller $w$) halo concentrations. As we show
below, a model with $w\approx -0.8$ and $\sige\approx 0.8$ can
describe the X-ray data in the cluster regime.

\subsection{Only Halos with $M>10^{14}\msun$}
\label{clusters}
\begin{figure}[t]
\centerline{\includegraphics[height=0.35\textheight]{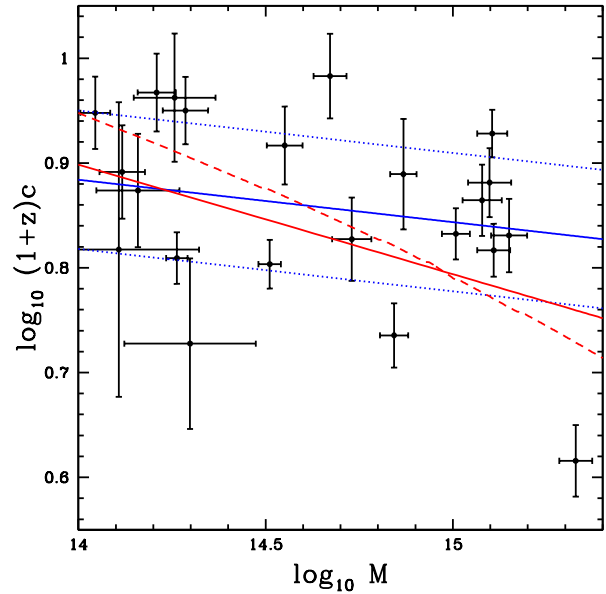}}
\caption{\label{fig.cm.clusters} 
Results of fitting eqn.\ \ref{eqn.cm} in log space only for halos with
$M>10^{14}\msun$. Displayed are the best-fitting model (solid blue
line) and the $1\sigma$ intrinsic scatter (dotted blue
lines). Also shown is the predicted relation for the
\lone\ model from cosmological simulations. The (solid red) line is
the power-law fit obtained by D04 while the (dashed red) line is the
B01 model with $F=0.001$ and $K=3.5$ found by D04 to best match the
simulated clusters. Note the D04 model is converted to our definition
of the virial radius (\S \ref{models}).}
\end{figure}

\begin{figure*}[t]
\parbox{0.49\textwidth}{
\centerline{\includegraphics[height=0.35\textheight]{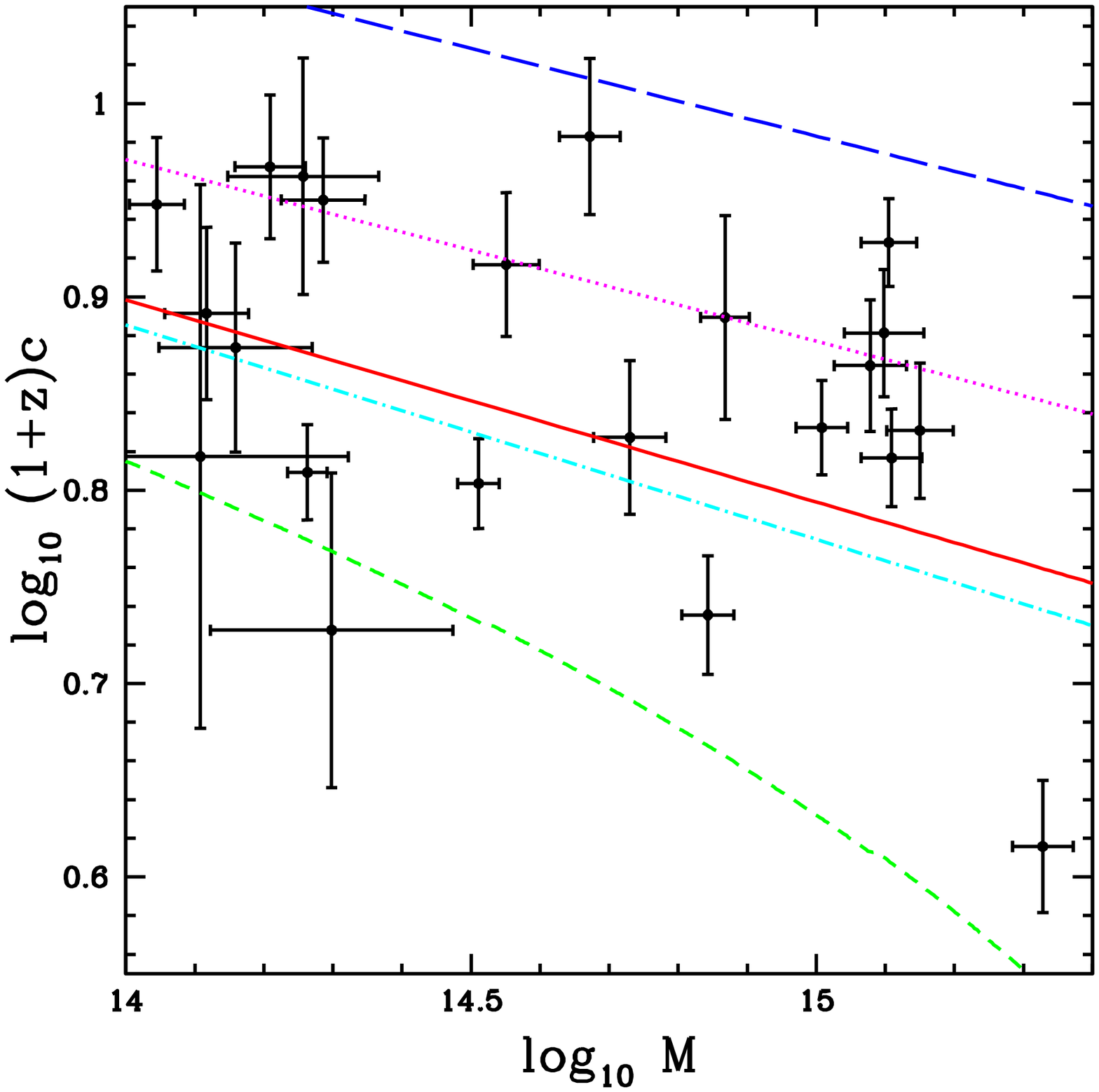}}}
\parbox{0.49\textwidth}{
\centerline{\includegraphics[height=0.35\textheight]{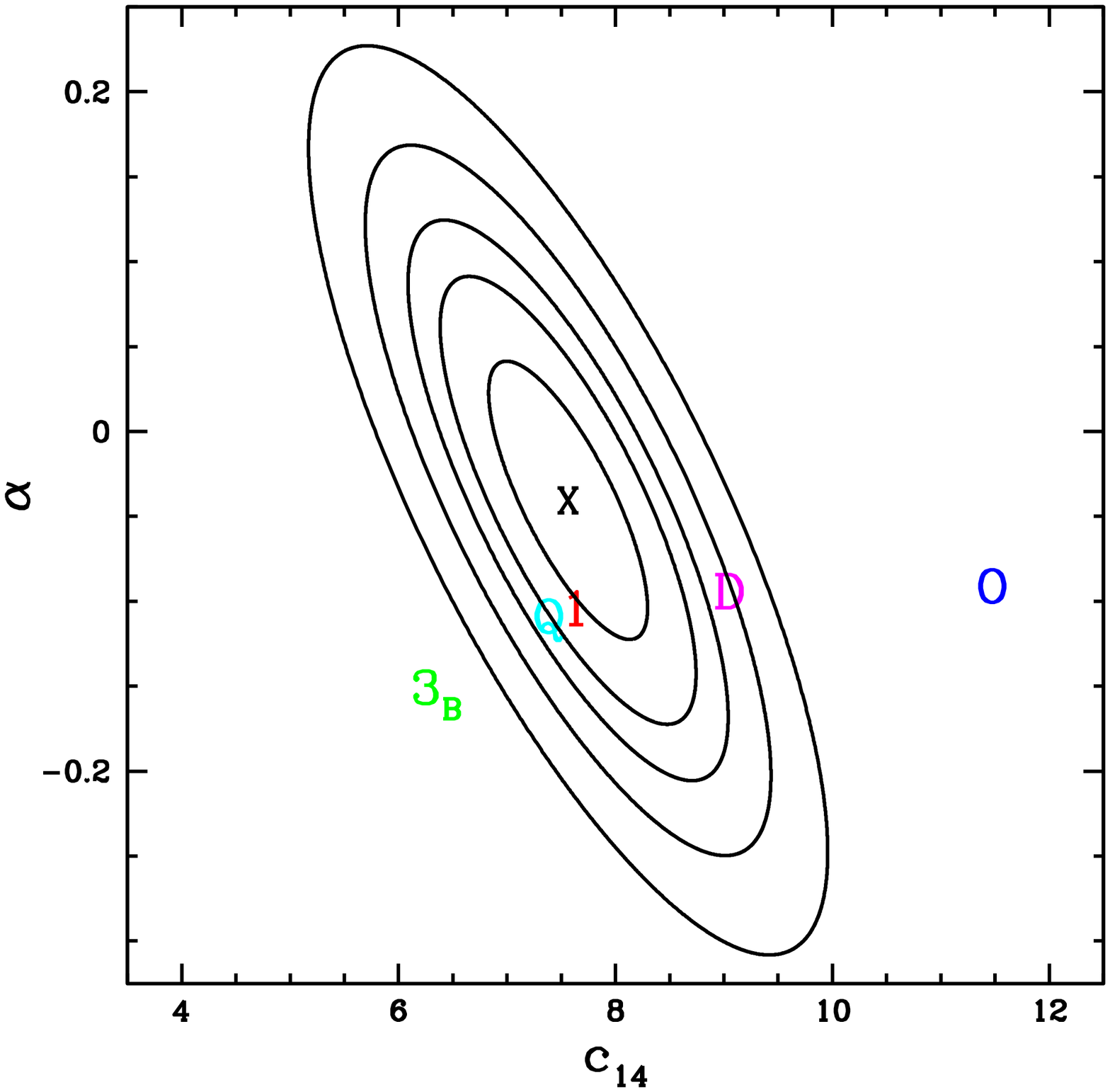}}}
\caption{\label{fig.errors.clusters} 
({\sl Left panel}) $c$-$M$ relations for various CDM models obtained
by D04 for galaxy clusters (see Table \ref{tab.models.cm}). The solid
(red) line is \lone, the dashed (blue) line is OCDM, the dotted
(magenta) line is DECDM, and the dot-dashed (cyan) line is
QCDM. Finally, the short dashed (green) line is the
\lthree\ model computed using the B01 prescription with $K=3.5$. 
({\sl Right panel}) Best-fitting value (X) and confidence contours
(68\%, 95\%, 99\%, 99.9\%, 99.99\%) for $\alpha$ and $c_{14}$ obtained
by fitting only systems with $M>10^{14}\msun$.  Also shown are the
results for the models plotted in the left panel: \lone\ (1), OCDM (O),
DECDM (D), QCDM (Q), and \lthree\ ($3_{\rm B}$).}
\end{figure*}
As discussed in \S \ref{models}, D04 provide results of fitting eqn.\
[\ref{eqn.cm}] to a suite of CDM models for massive clusters,
including an open model and multiple dark energy models.  We may
compare these theoretical predictions directly to the power-law fit to
the X-ray data with the following considerations. Firstly, we convert
D04's results to our definition of the virial radius (see \S
\ref{models}). Secondly, since D04 analyze only massive clusters
($M>4\times 10^{14}\msun$, $h=0.7$ and converted to $\Delta=101.1$), we
restrict our analysis to high-mass systems as well.  In order to allow
more precise constraints on the power-law fit we consider clusters
down to a somewhat smaller mass limit, $M>1\times 10^{14}\msun$. (As
shown below, this choice is justified since there is no obvious trend
in the $c-M$ relation over this mass range.)  Finally, we need to
consider the biases ($\approx 10\%$ higher $c$, smaller scatter in
$\logten\ c$) for relaxed, early forming systems as done for the whole
sample in the previous section. However, the systematic increase in
$c$ may be less since we are considering the most massive systems
\citep{wang06a}.

In Figure \ref{fig.cm.clusters} we show the result of fitting eqn.\
[\ref{eqn.cm}] to the high-mass sample, for which we obtain the
following parameter constraints, $\alpha = -0.04\pm 0.05$ and $c_{14}
= 7.6\pm 0.5$. These parameters are consistent with a constant $c-M$
relation, as well as profiles that both decrease and increase with
increasing $M$, and agree well with similar fits obtained in this mass
range from previous X-ray studies with either \xmm\ \citep{poin05a} or
\chandra\ \citep{vikh06a}. Since the power-law fit is quite consistent
with $\alpha=0$, it follows that it is the lower mass systems
($<10^{14}\msun$) which require $\alpha < 0$ found when fitting the
whole sample in the previous section \citep[see][]{gast06a}.  (Note
that the power-law parameters inferred in the previous section for the
full sample lies on the 99\% confidence contour obtained for the fit
to the high-mass subsample -- see Figure
\ref{fig.errors.clusters}.)

We measure an intrinsic scatter in $\logten\ (1+z)c$ of $0.07\pm 0.01$
which is somewhat less than the value of 0.10 obtained for the entire
sample. D04 also obtain 0.10 for the intrinsic scatter (after
accounting for a factor of 2.3 converting $\ln$ to \logten) for all of
their CDM clusters. The smaller scatter we measure presumably can be
explained because the X-ray sample preferentially contains relaxed,
early forming systems.

For comparison we plot two versions of the \lone\ model in Figure
\ref{fig.cm.clusters}. The first version we show is the power-law fit obtained by
D04 converted to our definition of the virial radius (see \S
\ref{models}). The second version arises from using the B01 model with
$K=3.5$, which D04 found to best represent their high-mass cluster
simulations. Both models are very similar and provide a reasonable
description of the X-ray data, though the B01 representation is slightly
steeper.

In the left panel of Figure \ref{fig.errors.clusters} we plot the
power-laws obtained by D04 for the CDM models as well as the B01
representation of the \lthree\ model with $K=3.5$.  The values of
$\alpha$ and $c_{14}$ for all of these models (Table
\ref{tab.models.cm}) are plotted in the right
panel of Figure \ref{fig.errors.clusters} along with the error
contours derived from the power-law fit to the X-ray data. As done in
the previous section, the $\alpha$ and $c_{14}$ values for the B01
\lthree\ model were obtained using the $c-M$ slope near
$M=10^{14}h^{-1}\msun$.

We emphasize that the D04 parameters may be
compared directly to the results of our power-law fit to the X-ray
data, since the same model is fitted over nearly the same mass
range. Since the B01 representation of the \lthree\ model is very
nearly a power-law over the range being investigated, a similarly
direct comparison is appropriate.

The results for the \lone\ and \lthree\ models obtained using only the
high-mass subsample corroborate those obtained from the full sample.
As noted above, the \lone\ model is an acceptable match to the X-ray
data; the parameters lie near the 68\% confidence contour in Figure
\ref{fig.errors.clusters}, even if $c$ is increased to correct for
early forming halos.  In contrast, the \lthree\ model is very
inconsistent with the high mass data, with parameters lying outside
the 99.99\% contour.  Even if we apply the full 10\% correction to
$c_{14}$ appropriate for the most relaxed, early forming systems, then
the model merely moves on top of the 99.99\% contour. However, as
noted above, it is expected that this correction is less for the
highest mass systems. Hence, the \lthree\ model is also rejected using
only the most massive clusters.

The $c-M$ data for high-mass clusters clearly exclude the OCDM
model. The concentrations lie systematically above the data and the
\lone\ model. The latter is expected because structures form earlier
in the OCDM model. In principle, the OCDM model may be brought into
acceptable agreement with the X-ray data by lowering \sige.  We
estimate the effect of lowering \sige\ using the \lone\ B01 model. For
example, lowering $\sige$ from 0.90 to 0.70 lowers $c_{14}$ by
25\%. This would move the OCDM model to the right in Figure
\ref{fig.errors.clusters} on to the 99\% confidence
contour. Considering any of the 10\% systematic bias expected for
early forming systems would increase the discrepancy. (Note that for
$\sige=0.76$, appropriate for the \lthree\ model, the OCDM model would
be rejected at the $\approx 99.9\%$ confidence level without additional
consideration of early formation bias.)

Studies of the abundances of galaxy clusters using weak gravitational
lensing typically find \sige\ consistent with 0.9
\citep[e.g.,][]{hoek02a,vanw05a,jarv03a}. The optical study of
\citet{rine06b} uses the caustic technique to measure cluster abundances
and also finds $\sige\approx 0.9$ with a lower limit of 0.72 (95\%
conf.). However, X-ray studies of cluster abundances show a large
variation in \sige, with values ranging from 0.7-0.9 \citep[see][and
references therein]{arna05a}. The X-ray studies use a simple
conversion between mass and global temperature or mass and global
X-ray luminosity ($\lx$).  A recent study by
\citet{stan06a} argues that considering reasonable errors in the $M-\lx$
conversion allows agreement of cluster abundances with the recent WMAP
parameters with $\sige=0.85$ and $\omegam=0.24$ -- though see
\citet{reip06a}.  We take $\sige=0.7$ to be a conservative lower
limit established by the abundances of galaxy clusters.

Therefore, an open CDM model with $\omegam\approx 0.3$ is ruled out
($>99\%$ confidence) from joint consideration of only the X-ray $c-M$
relation and cluster abundances. Since $\sige$ and $\omegam$ are
tightly coupled from cluster abundance studies, and all the studies
cited above give $\sige\approx 0.45$ for $\omegam\approx 1$, using the
X-ray $c-M$ data we find that we can also exclude CDM models with
$\omegam\approx 1$ at high significance ($>99.9\%$ confidence).  For
this comparison we have employed the entire sample of 39 systems and
have set $h\approx 0.5$ for the $\omegam\approx 1$ models to better
satisfy cosmic age constraints from stellar populations in globular
clusters \citep{chab98a}. Consequently, the combination of the X-ray
$c-M$ relation and cluster abundances, assisted by constraints on the
cosmic age, provides novel evidence for a flat, low-$\omegam$ universe
with dark energy using observations only in the local ($z\ll 1$)
universe.

Finally, we consider the alternative dark energy models, DECDM and
QCDM. The DECDM model lies systematically above the \lone\ model and
deviates from the power-law fit to the X-ray data at the 99.9\%
confidence level. This behavior is expected since $w=-0.6$ implies
halos form earlier compared to the $w=-1$ \lone\ model \citep[see
also][]{kuhl05a}. By lowering \sige\ it should be possible to bring
the DECDM model into good agreement with the data. To illustrate this
trade-off between $w$ and \sige, we consider the QCDM model which has
$w\approx -0.8$ and $\sige=0.82$. As shown in Figure
\ref{fig.errors.clusters} the QCDM model matches the X-ray results
nearly as well as the \lone\ model.  In fact, a mild increase in
$c_{14}$ owing to effect from selecting relaxed, early forming halos
would produce even better agreement.

Our analysis indicates that a model similar to QCDM is able to satisfy
the most recent constraints from other cosmological observations since
$w\approx -0.8$ is marginally consistent with the constraints imposed
by supernovas \citep[$w=1.023 \pm 0.090\, \rm (stat) \pm 0.054\,
(sys)$, ][]{asti06a}, and $\sige=0.82$ is also marginally consistent
with the 3-yr WMAP results. Put another way, if the low \sige\ value
of \lthree\ is correct, then the X-ray $c-M$ data imply $-1 < w\approx
-0.8$.

\section{Systematic Errors}
\label{sys}

When comparing to theoretical predictions of the $c-M$ relation we
have considered the small biases in the mean concentration level and
scatter expected for the highly relaxed, early forming systems
selected for our study.  We found that the \lone\ model is a good
representation of the X-ray data provided this bias for relaxed systems
applies, especially for the low-mass range of our sample
($M<10^{14}\msun$). However, the \lthree\ model does not agree with
the $c-M$ data, even considering this bias. 

Could other sources of systematic error seriously affect these
conclusions? For 24 of 39 systems in Table \ref{tab.data} we have
performed a detailed investigation of systematic errors on $c$ and $M$
\citep{hump06b,zapp06a,gast06a}. Generally, we find that the estimated
systematic error is less than the statistical error, particularly for
the systems of lowest mass.  For the higher mass systems, the errors
are usually comparable. But from consideration of all of these systems
we recognize no obvious trend that would systematically shift the
observed $c-M$ relation in one direction. 

Although we do not believe systematic errors associated with the X-ray
data analysis compromise the conclusions of our present investigation,
here we list the most important issues to be resolved in future
studies that seek to use the X-ray $c-M$ relation for precision
constraints on cosmological models.  Below in \S \ref{he} we also
consider the impact on our present analysis of systematic errors
resulting from the hydrostatic equilibrium approximation.

(Note we verified that our analysis is insensitive to the definition
of the virial radius. An interesting example is to compare to results
obtained for $\Delta=2500$, because within this radius all of the
systems in our study possess good X-ray constraints. When defining the
virial radius to correspond to $\Delta=2500$ for every system, we
arrive at the same conclusions obtained in \S \ref{results} but with
correspondingly different $\alpha$ and $c_{14}$ values. The
predictions of the cosmological models were also converted to
$\Delta=2500$ for this comparison.)

\subsection{Early Formation Bias}
Accounting for the biases associated with preferentially observing the
most relaxed, early forming systems is required for precision
constraints on cosmology. While \cite{wech02a} found a nearly
one-to-one correlation between halo ``formation epoch'' and
concentration, it is not obvious how this theoretically motivated
parameter should connect with the dynamical state of a real cluster.
More direct quantifiers (e.g. rejecting cases with recent major
mergers and disturbed profiles) suggest that ``relaxed'' halos should
have $\approx 10\%$ larger average $c$ and smaller scatter. For
massive systems $(M\ga (\rm few)\times 10^{14}\msun$), the outlook for
analyzing a sample with a well-determined, observational selection
function is excellent, because there exist several well-defined
catalogs of the brightest, most massive clusters in X-rays
\citep[e.g.,][]{reip02a}, of which many of the systems have been
observed either with \chandra\ or \xmm. For lower masses the number of
X-ray catalogs of complete samples is small, and the number with good
coverage with \chandra\ or \xmm\ observations is even smaller. For
this purpose we have scheduled observations of a complete, X-ray
flux-limited sample of 15 systems in the approximate mass range
$10^{13}-10^{14}\,\msun$ with \chandra.

\subsection{Hydrostatic Equilibrium}
\label{he}

The determination of the mass distribution from X-ray observations
requires that hydrostatic equilibrium is a suitable
approximation. This approximation has been tested for massive clusters
by comparing to results obtained from gravitational lensing
\citep[e.g.,][and references therein]{buot03c}. There is very good
agreement between the methods, especially outside the inner
cores. This agreement is especially encouraging since some of the
clusters are manifestly not completely relaxed \citep[e.g., A2390,
][]{alle01d}.  At the low-mass end, good agreement between X-ray mass
measurements with stellar dynamics in elliptical galaxies provides
further indication that the hydrostatic equilibrium approximation is
quite accurate for obviously relaxed systems \citep{hump06b,brid06a}.

For over ten years cosmological hydrodynamical simulations have found
that the hydrostatic equilibrium approximation is quite accurate in
massive galaxy clusters
\citep[e.g.,][]{tsai94a,buot95a,evra96a,math99a}.  The most recent
studies conclude that X-ray mass estimates of the most massive,
relaxed clusters should typically underestimate the mass by a small
amount ($\approx 10\%$) because of turbulent pressure in the hot gas,
with less of an effect in lower mass systems
\citep[e.g.,][]{dola05a,rasi06a,naga06a}.

The systems analyzed in our present investigation were selected to be
highly relaxed -- as indicated by regularly shaped X-ray image
morphology. This is reinforced by our analysis of the X-ray $c-M$
relation (\S \ref{results}), especially by the small intrinsic scatter
in the concentrations which is a robust prediction of CDM models
\citep{jing00a,dola04a,wech02a,macc06a}.

Underestimating the virial masses of our relaxed clusters by $10\%$,
as suggested by CDM simulations, would indicate that our measurement
of $c_{14}$ should be raised by a factor $1.1^{-\alpha}\approx 1.02$,
using $\alpha=-0.17$.  In addition, since $c\propto
(\mvir)^{1/3}/r_s$, where $r_s$ is the NFW scale radius, the total
increase in $c_{14}$ should be $\approx 5\%$, provided the estimate of
$r_s$ is unaffected by the presence of turbulent pressure.  This
systematic error, if real, would have the effect of increasing the
discrepancies between the X-ray $c-M$ relation and (most) CDM
models. However, essentially all of conclusions in \S \ref{results}
remain unchanged if we consider the bias for relaxed early forming
systems to be 15\% rather than 10\%, within the range obtained by
current simulations \citep{jing00a,wech02a,macc06a}.  The exceptions
are the OCDM and DECDM models (\S \ref{clusters}, see Fig.\
\ref{fig.errors.clusters}), for which the $\approx 5\%$ boost in
$c_{14}$ would bring the models into slightly better agreement with
the data. However, the OCDM model is still rejected at $>99.99\%$
confidence, and the $\omegam\approx 0.3$ open CDM models are still
excluded ($>99\%$ conf.), the latter provided that at least a 5\% bias
(of the expected $\approx 10\%$) for relaxed, early forming systems
applies. We conclude, therefore, that the level of systematic error
suggested by CDM simulations to affect the X-ray mass measurements
does not change the conclusions of our present study.

\subsection{Semi-Analytic Model Predictions of $c-M$ relation}
\label{fk}

In order to use the X-ray $c-M$ relation for precision constraints on
cosmological parameters, it is necessary to be able to predict halo
concentrations produced in different cosmological models with high
precision. This must be achieved with a semi-analytic procedure,
because it is not feasible to resort to N-body simulations to fully
investigate parameter space for obtaining confidence regions.  The
procedure proposed by B01 is currently one of the most promising
models of this kind.  With just two parameters $F$ and $K$ (see \S
\ref{models}) it can reproduce the results of CDM N-body simulations
with ranges of power spectrum shapes, $\sige$ normalizations, matter
content, and dark energy variables $w$ (B01, \citealt{kuhl05a}).

Recently, \citet{macc06a} show that the normalizations of the B01
models (i.e., $K$ values) obtained by different investigators
analyzing halos covering approximately the same mass range ($\la
10^{13}h^{-1}\msun$) can differ by 10-20\%.  
For halos with $M \approx
10^{13}- 10^{14} h^{-1}\msun$, the halos of
\citet{macc06a} prefer $F=0.001$ and $K=2.6$ while those
\citet{tasi04a} prefer $K=3.1$.  The ``cluster'' halos ($\ga 3\times
10^{14}\msun$) studied by D04 suggest a higher normalization
$K=3.5$.  Because so few
studies have investigated the high-mass halo regime, it is unclear
whether these differences reflect numerical details in the simulations
or demand a revision of the B01 model. More studies of the
concentrations of the highest mass halos are very much needed.

In this context it is important to note that \citet{shaw06a} have
recently studied the $c-M$ relation of massive halos, where most of
their halos have $M\approx 1\times 10^{14}\msun$. They obtain
power-law fit parameters $\alpha=-0.12\pm 0.03$ and $c_{14}=8.30 \pm
0.04$ for the model \lone\ but with $\sige=0.95$. Converting their
results to $\sige=0.90$ yields a 6\% reduction in the concentration,
$c_{14}=7.81 \pm 0.04$, which is just 2\% larger than the D04 value
quoted in Table \ref{tab.models.cm}. We consider this quite good
agreement, considering the differences between theoretical studies
noted above for lower mass halos.

Finally, we mention that if we adopt the lowest normalization quoted
in the literature ($F=0.001$, $K=2.6$, \citealt{macc06a}) for our
entire mass range, then in order for the \lone\ model to match the
X-ray data, it is necessary to boost the concentrations produced in
the \lone\ model by another 16\% over the $10\%$ attributed to the
preferential selection of relaxed, early forming systems. This 16\%
increase can be achieved by increasing either \sige ($\approx 1.0$),
$w$ ($\approx -0.8$), or both.  We note that the 99\% upper limit
derived in \S \ref{all} would increase \sige\ from 1.07 to 1.15 if
$K=2.6$.

\subsection{Gas Physics \& Adiabatic Contraction}

The published CDM predictions for the $c-M$ relation we have
considered in this paper are all derived from dissipationless N-body
simulations containing only dark matter. While the details of
gasdynamics should be unimportant for massive systems ($M\ga
10^{14}\msun$), the effects of dissipation and feedback from star
formation and AGN likely influence the dark matter profile inferred
from observations of lower mass halos. It is noteworthy that the
systems with $M\sim 10^{13}\msun$ show the largest deviations from the
\lone\ model (\S \ref{all}). For most of these the \chandra\ data
require a significant contribution of stellar mass from the central
galaxy \citep{hump06b,gast06a}. However, allowing for adiabatic
contraction \citep[e.g.,][]{blum86a,gned04a} of the dark matter
profile in most cases degrades the fits, suggesting a more complex
interplay between the baryons and the dark matter
\citep[e.g.,][]{loeb03a,elza04a,dutt06a}.  Some cosmological
simulations with gas predict a small, but significant, systematic
increase ($\approx 3\%$) in the concentration of the total mass
\citep{lin06a}, but the inability of CDM simulations to reproduce
observed X-ray temperature and density profiles within 50-100~kpc of
cluster centers \citep[e.g.,][]{lewi00a,muan02a,borg04a} makes it
difficult to interpret the reliability of such results.  Future
precision cosmology studies which aim to use the X-ray $c-M$ relation
for low-mass systems ($M\la 10^{13}\msun$) will require better
understanding of the influence of the central galaxy on the inferred
dark matter distribution.

\section{Conclusions}
\label{conc}

We present the concentration ($c$)-virial mass ($M$) relation of 39
galaxy systems ranging in mass from individual early-type galaxies up
to the most massive galaxy clusters, $(0.06-20) \times
10^{14}\msun$. We selected for analysis the most relaxed systems
possessing the highest quality data currently available in the
\chandra\ and \xmm\ public data archives. Measurements for 24
systems were taken from our recent work
\citep{hump06b,gast06a,zapp06a} which populate the lower mass portion
of the sample ($M\la 10^{14}\msun$).  We obtained results for 15
massive galaxy clusters in our sample from the studies by
\citet{poin05a} and \citet{vikh06a}.

Our principal objective is to measure the $c-M$ relation accurately
from galaxy to cluster scales and determine whether $c$ decreases with
increasing $M$ as generally predicted by CDM models. However, we also
use the $c-M$ relation to provide an initial demonstration of the
ability of the $c-M$ relation to constrain cosmological parameters,
which also serves to highlight key sources of systematic error -- both
in the theoretical models and the observations.

We parameterize the X-ray $c-M$ relation using a simple power-law
model.  The best estimates of the parameters -- the slope ($\alpha$)
and normalization ($c_{14}$) evaluated at
$M=M_{14}=10^{14}h^{-1}\msun$ -- were obtained via linear regression
in log space taking into account the uncertainties on both $c$ and
$M$. We employed the BCES method of \citet{akri96} for this analysis.

Fitting the power-law model to the entire sample yields
$\alpha=-0.172\pm 0.026$ and $c_{14}=9.0\pm 0.4$ (quoted errors
$1\sigma$). The slope $\alpha$ is negative and inconsistent with 0 at
$6.6\sigma$.  The previous studies of galaxy clusters ($M\ga
10^{14}\msun$) with \chandra\ and \xmm\ by \citet{poin05a} and
\citet{vikh06a} did not place strong constraints on $\alpha$, and were
quite consistent with $\alpha=0$; i.e., it is the lower mass galaxy
groups that require $\alpha < 0$ \citep[see][]{gast06a}. Recent
optical studies of the $c-M$ relation in the group-cluster regime (see
\S \ref{intro}) also do not place strong constraints on the slope and
are very consistent with $\alpha = 0$. Our analysis, therefore,
provides crucial evidence that $c$ decreases with increasing $M$, as
expected in CDM models (\citealt{nfw,jing00a}; B01; D04;
\citealt{macc06a}).

We compare the X-ray data to the \lone\ model (with $\omegam=0.3$,
$\sige=0.9$, $n_s=1$), the ``concordance model'' effectively
representing the combined constraints from the first year of WMAP CMB
data, supernovae, and galaxy surveys \citep[e.g.,][]{sper03a}.  We
judge the median $c-M$ relation of the \lone\ model to be consistent
with the empirical power-law fit provided the X-ray sample consists of
the most relaxed, early forming systems, for which a systematic
increase in the concentrations of $\approx 10\%$ is expected
\citep{jing00a,wech02a,macc06a}. We measure an intrinsic scatter in
$\logten\ (1+z)c$ of $0.102\pm 0.004$ in excellent agreement with the
prediction of CDM simulations for the most relaxed, early forming
systems.  The amount of scatter is a robust prediction of CDM variants
and provides additional evidence that our sample comprises the most
relaxed systems that are best suited for X-ray mass measurements
requiring approximate hydrostatic equilibrium.

The X-ray $c-M$ relation places interesting constraints on
\sige. Within the context of the concordance model noted above, the
$c-M$ relation requires $0.76<\sige<1.07$ (99\% conf.), assuming a
10\% upward bias in the concentrations for early forming systems.
This confidence range is conservative as explained in \S \ref{all}.

Next we compare the X-ray data to the \lthree\ model (with
$\omegam=0.24$, $h=0.73$, $\sige=0.76$, $n_s=0.96$) which effectively
updates the \lone\ model using the analysis of the third year of WMAP
data \citep{sper06a}. The X-ray $c-M$ relation rejects the \lthree\
model at the 99.99\% confidence level. The primary reason for the poor
performance of the \lthree\ model is the low value of
\sige, but the lower value of \omegam\ and the tilt of the power
spectrum also contribute to the poor fit.  For this comparison we have
assumed a uniform bias for relaxed, early forming halos of $\approx
10\%$, as suggested by numerical simulations. This bias would have to
be $\approx 50\%$ to bring the \lthree\ model into agreement with the
X-ray data.

While the early-type galaxy and group-cluster mass halos $(\ga
10^{13}\msun)$ studied here apparently prefer slightly higher
concentrations than predicted for typical halos in the concordance
$\Lambda$CDM model, the opposite is true for late-type galaxies
\citep[e.g.,][]{alam02a,bosc03a,dutton05,kuz06,gnedin06}.  
This may indicate that a selection/formation-time bias operates across
the galaxy type spectrum, with late-type galaxies inhabiting the
low-concentration tail of the distribution \citep[though
see][]{napo04,napo05}.  

Since D04 provide results of power-law fits to the $c-M$ relations of
massive clusters formed in a variety of CDM models, including an open
model and several dark energy models, we analyze separately the X-ray
$c-M$ relation for the 22 systems in our sample with
$M>10^{14}\msun$. In this mass range we obtain $c_{14}$ and
$\alpha\approx 0$ values consistent with those inferred from previous
\chandra\ and \xmm\ studies in the cluster mass regime noted above
\citep{poin05a,vikh06a}. As also found  for the entire sample,
the $c-M$ relation of the high-mass subsample is consistent with the
\lone\ model and very inconsistent with the \lthree\ model.

However, an open model with $\omegam\approx 0.3$ is ruled out ($>99\%$
confidence) from joint consideration of only the X-ray $c-M$ relation
and published constraints on \sige\ from the analysis of the
abundances of galaxy clusters ($\sige > 0.70$, see \S
\ref{clusters}). Since cluster abundance studies also find
$\sige\approx 0.45$ if $\omegam\approx 1$, using the X-ray $c-M$ data
we find that we can also reject CDM models with $\omegam\approx 1$ at
a high significance level ($>99.9\%$ confidence). Consequently, the
combination of the X-ray $c-M$ relation and cluster abundances (and
local constraints on the age of the universe) provides novel evidence
for a flat, low-$\omegam$ universe with dark energy using observations
only in the local ($z\ll 1$) universe.

If the values of \sige, \omegam, and $n_s$ of the \lthree\ model are
correct, agreement with the X-ray $c-M$ relation may be achieved by
increasing the dark energy equation of state parameter $w$. We find
that a quintessence model with $w\approx -0.8$ and $\sige=0.82$
performs nearly as well as the \lone\ model, and the larger value of
$w$ remains marginally consistent with supernova constraints
\citep{asti06a}.

Finally, we discuss key sources of systematic error associated with
both the X-ray measurements and theoretical models that need to be
addressed before the X-ray $c-M$ relation is suitable for precision
cosmology. In particular, if the virial masses are systematically
underestimated by $\sim 10\%$, as suggested by CDM simulations, then
we estimate that $c_{14}$ is increased by $\sim 5\%$, less than the
$\approx 10\%$ increase expected for selecting relaxed, early forming
systems. This level of systematic error does not change the
conclusions of our present study (see \S \ref{he}), but it will be
important for future studies of precision cosmology.

\acknowledgements 
We thank A.\ Cooray for discussions and comments on the manuscript.
D.A.B., P.J.H., and F.G.\ gratefully acknowledge partial support from
NASA grants NNG04GE76G, issued through the Office of Space Sciences
Long-Term Space Astrophysics program, and NAG5-13059, issued through
the Office of Space Science Astrophysics Data Program.  Partial
support for this work was also provided by NASA through Chandra Award
Numbers GO4-5139X and GO6-7118X issued by the Chandra X-ray
Observatory Center, which is operated by the Smithsonian Astrophysical
Observatory for and on behalf of NASA under contract NAS8-03060. We
also are grateful for partial support from NASA-XMM grants,
NAG5-13643, NAG5-13693, NNG04GL06G, and NNG05GL02G.


\end{document}